\documentclass[conference]{IEEEtran}
\usepackage[cmex10]{amsmath}
\usepackage[cmex10]{amsmath}

\usepackage{ifpdf}
\usepackage{xurl}
\usepackage{algorithm}
\usepackage{algpseudocode}
\usepackage{xcolor}
\usepackage{url}
\ifpdf
\usepackage[pdftex]{graphicx}
\else
\usepackage[dvips]{graphicx}
\fi
\usepackage{tabularx}
\usepackage{multirow}
\graphicspath{{graphics/png/}}
\DeclareGraphicsExtensions{.png, .eps, .pdf}
\usepackage{float} 
\usepackage{subfigure}
\usepackage{comment}
\usepackage{bm}
\usepackage{amsmath,amsthm,amsfonts,amssymb}
\usepackage{pifont}
\usepackage[noadjust]{cite}
\usepackage{nomencl}
\usepackage{hyperref}

\theoremstyle{definition}

\newcommand\Tstrut{\rule{0pt}{2.6ex}}         

\IEEEoverridecommandlockouts
\begin{document}
%
\title{A WECC-based Model for Simulating Two-stage Market Clearing with High-temporal-resolution}

\author{\IEEEauthorblockN{Ningkun Zheng, Bolun Xu}
\IEEEauthorblockA{\textit{Earth and Environmental Engineering} \\
\textit{Columbia University}\\
New York, USA\\
\{nam2252, nz2343, nq2176, bx2177\}@columbia.edu}
}


\maketitle

\begin{abstract}
This paper presents a new open-source model for simulating two-stage market clearing based on the Western Electricity Coordinating Council Anchor Data Set. We model accurate two-stage market clearing with day-ahead unit commitment at hourly resolution and real-time economic dispatch with five-minute resolution. Both day-ahead unit commitment and real-time economic dispatch can incorporate look-ahead rolling horizons. The model includes seven market regions and a full year of data, detailing 2,403 individual generation assets across diverse energy sources. The year-long simulation demonstrates the capability of our model to closely reflect the generation and price patterns of the California ISO. Our sensitivity analysis revealed that extending the ED look-ahead horizon reduces system costs by up to 0.12\%. We expect this new system model to fulfill the needs of conducting electricity market analysis at finer time granularity for market designs and emerging technology integration. While we focus on the western interconnection, the model serves as a base to simulate other two-stage clearing market locations.


\end{abstract}

\begin{keywords}
Economic dispatch, energy storage, power system economics, unit commitments.
\end{keywords}

\IEEEpeerreviewmaketitle

\section{Introduction}
The electric power sector has seen significant advancements in emerging technologies aimed at reducing greenhouse gas emissions and supporting decarbonization. Rapid growth in renewable energy and battery storage capacity reflects this shift, particularly in regions like Texas and California with ambitious targets for 2045. These developments push electricity markets to adapt to the operational characteristics of renewable and other emerging technologies.

Several open-source test systems have been developed to evaluate and benchmark power systems, including the IEEE test systems. However, many rely on outdated or synthetic data, limiting their applicability to today’s challenges~\cite{10146422}. Efforts to incorporate more realistic data for test systems have led to models that address dynamic analysis, power flow (PF) and optimal power flow (OPF), and market simulation. For example, Tesfatsion et al. and Battula et al. have designed test systems focused on market simulations for regions like ISO New England and ERCOT, but gaps remain in high-temporal-resolution modeling for real-time applications~\cite{krishnamurthy20158, battula_ercot_2020}.

Benchmarking emerging technologies in power systems has become essential, particularly as technologies like energy storage, demand response, and hydrogen production add complexity to real-time operations~\cite{xu2017scalable, oldewurtel2013framework, sun2021integration}. Studies emphasize the importance of high-resolution market models to reflect operational dynamics, especially with intermittent and flexible resources~\cite{ela2013impacts, menati2023high}. High-fidelity test systems with detailed dispatch and look-ahead capabilities are crucial to evaluate the economic and operational impacts of these technologies, making it possible to explore real-time balancing, pricing, and resource adequacy in a rapidly evolving grid landscape.

In this paper, we introduce an open-source test system designed for high-temporal-resolution real-time market simulations with day-ahead and real-time market capabilities. Our key contributions include:

\begin{itemize}
    \item We develop a WECC-based model for two-stage wholesale market simulation, utilizing publicly accessible datasets and designed in the Julia language. This model offers high-temporal resolution and multi-interval look-ahead horizon functions in both DAM UC and RTM ED with tractable computational cost, benefiting from Julia's computational efficiency. 
    \item The yearly simulation using our proposed model closely emulating the generation and price trends observed in the California ISO (CAISO). Such accuracy is invaluable for modeling price-sensitive resources, like energy storage systems and demand response.
    \item A sensitivity analysis with varied look-ahead horizons in DAM UC and RTM ED highlights the advantages of the look-ahead window, especially in the effective management of emerging technologies.
\end{itemize}

Though focused on CAISO, the model is adaptable to other markets, providing a versatile tool for studying the integration and impact of emerging technologies in wholesale markets.

\section{Test System Development}
We outline the data collection and processing approach from open-source data. We first discuss the simplified WECC system, including regions, network, generation resources, renewable and load profiles~-~developed from the WECC 2032 ADS PCM (Anchor Data Set Production Cost Model) Public Data, reduced WECC 240-bus system~\cite{price2011reduced}, and CAISO Demand Forecast Data. This two-stage wholesale electricity market model comprises an hourly-resolution DAM and intra-hourly RTM, both with multi-period optimization and look-ahead window functionality.

\subsection{Data Processing}
WECC ADS data, sourced from multiple US, Canadian, and Mexican entities, represents projected network, generation resources, and loads over a 10-year planning horizon. This ensures a representation of the present and projected network, generation resources, and loads over a ten-year planning horizon that closely reflects real-world conditions. In the publicly available version of WECC ADS, transmission data is withheld due to security concerns. Consequently, we utilize the reduced WECC 240-bus model (WECC-240) to derive transmission data. While WECC ADS was initially designed for hourly-resolution production cost models and doesn't provide high-temporal-resolution renewable and load profiles, we generate real-time profiles based on CAISO historical data.

\subsubsection{Simplified Regions and Transmission}

\begin{figure}
    \centering
    \includegraphics[trim = 5mm 5mm 70mm 10mm, clip, width=.60\columnwidth]{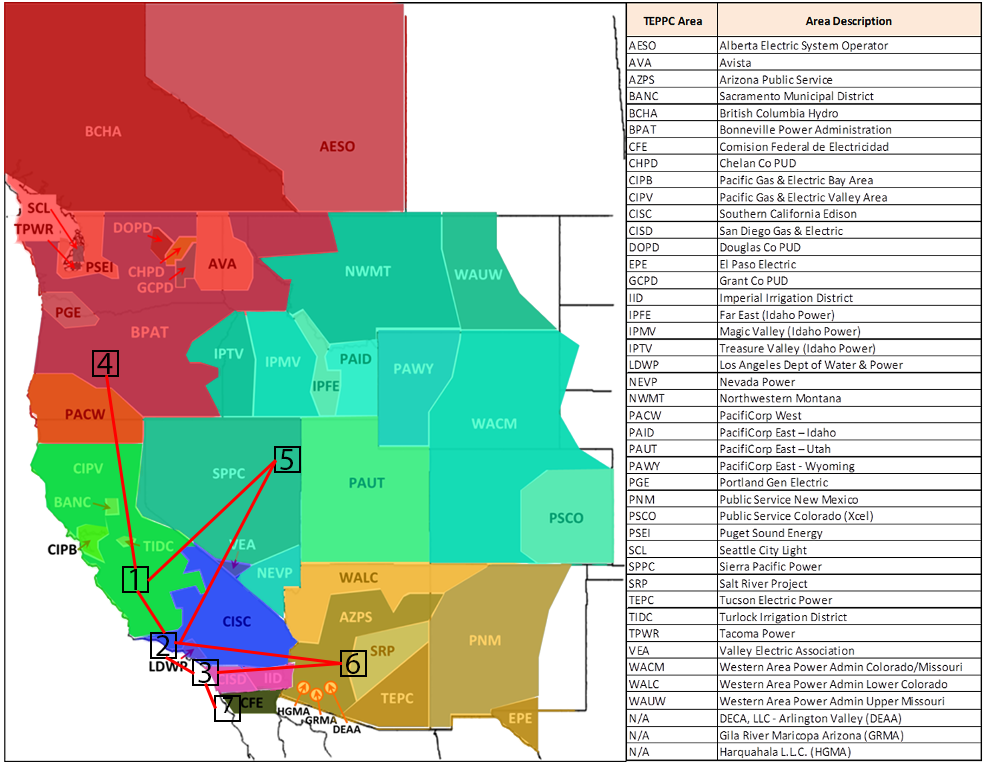}
    \caption{Reduced and modified regions and network for WECC.}
    \label{fig:ZoneMap}
        \vspace{-1em}
\end{figure}

The transmission system is simplified to seven regions, as illustrated in Fig.~\ref{fig:ZoneMap}, a choice driven by the unavailability of detailed transmission data in the open-source ADS. Regional transmission limitations are still represented using selected lines from the WECC-240 and capacity limits from EIA-930 data, ensuring a realistic portrayal of transmission constraints despite the simplified model. Our approach computed line impedances within the WECC-240 using parallel impedance calculations on manually selected lines, only considering reactance. To better represent transmission congestion, we set transmission line capacity as the maximum hourly interchange data from neighboring balancing authorities in 2022, sourced from the EIA-930. Parameters of transmission lines are detailed in Table.~\ref{tab:Transmission}.

\begin{table}[ht]
\footnotesize
\caption{Transmission Line Parameters}
\centering{
\begin{tabular}{cccc}
\hline
\hline
\Tstrut
From Region & To Region
& Reactance X (p.u.)  & Capacity (MW)\\
\hline
1&2&0.0071&6,368\\
\hline
1&4&0.0034&3,500\\
\hline
1&5&0.0299&5,500\\
\hline
2&3&0.0287&888\\
\hline
2&5&0.0195&1,700\\
\hline
2&6&0.0436&3,200\\
\hline
3&6&0.0209&2,200\\
\hline
3&7&0.0270&1,776\\
\hline
\hline
\end{tabular}
}
\label{tab:Transmission}
    \vspace{-1em}
\end{table}

\subsubsection{Generation Resources Data}
Generation resource data is sourced from ADS, filtered to include only current existing resources: 45 coal, 925 natural gas, 6 nuclear, 167 biofuel, 123 geothermal, 669 solar PV, 351 wind, 1134 hydro, and 82 storage units. To capture unit commitment and dispatch effects, we model each unit individually for realistic market price simulation. For fast-start units, we use a piece-wise linear cost model in RTD, assigning costs from \$200 to \$400 per MW over 0-1000 MW capacity.

The total nameplate capacity of thermal generators, which includes geothermal and nuclear resources, stands at 115,320~MW. Table.~\ref{tab:RenewableCapacity} compares the installed renewable resources in CAISO from our processed data with the actual figures at the end of 2022~\cite{CAISOKeyStats2022}. Notably, the renewable penetration in the table closely aligns with real-world statistics.

\begin{table}[ht]
\footnotesize
\caption{Comparison of Test System Renewable Resources with Actual Installations in CAISO (End of 2022)}
\centering{
\begin{tabular}{ccccccc}
\hline
\hline
\Tstrut
&Wind&Solar&Hydro&Geo&Bio&Battery\\
\hline
Test System&6,485&16,007&10,077&903&2,583&4,775\\
\hline
Actual&7,950&15,967&12,281&798&1,597&4,614\\
\hline
\hline
\end{tabular}
}
\label{tab:RenewableCapacity}
    \vspace{-1em}
\end{table}

We model thermal generators using multi-segment piece-wise linear approximated heat rate curves. We directly sourced the maximum/minimum capacity, initial dispatch, segment increment capacity,  segment increment heat rate, minimum up/down time, ramp up/down rate, variable operation and maintenance cost, and fuel cost from the ADS. The start-up cost, no-load cost, and segment marginal cost are calculated as detailed below:
\begin{subequations}
\begin{align}
    \mathrm{SUC}_g &= \mathrm{SUC}_g^{fix} + \mathrm{FC}_g * \mathrm{SUF}_g \\
    \mathrm{NLC}_g &= \mathrm{FC}_g * \mathrm{FMin}_g\\
    \mathrm{MC}_{g,gs} &= \mathrm{FC}_g * \mathrm{HR}_{g,gs}
\end{align}
\end{subequations}
where $SUC_g$, $NLC_g$, and $MC_{g,gs}$ are start-up cost, no-load cost, and segment marginal cost of thermal generator $g$, respectively. $SUC_g^{fix}$ is the fixed start-up cost (\$), $FC_g$ is the fuel cost (\$/MMBtu), $SUF_g$ is the start-up fuel (MMBtu), $FMin_g$ is the minimum fuel input (MMBtu), $HR$ is the linearized heat rate for segment $gs$. To better match real bidding costs during peak loads in our simulation, we increased fuel prices by 20\%, mirroring the observed average bidding price rise in generators operating at peak load seasons~\cite{liu_open_2022}.

The ADS provides unit-wise max capacities and hourly profiles for wind and solar, while hydro profiles are based on 2018 generation and summarized by load areas. For hydropower plants without profiles, we use the nearest load area's data. Due to data limitations, we omit detailed water management, modeling hydro output to align with max hourly outputs in the ADS "Hydro-Given Schedule." For pumped hydro and battery storage, only max/min power capacities are available in the ADS. Their durations are set to 12 and 4 hours, with one-way efficiencies of 80\% and 90\%, respectively.

\subsubsection{WECC Load and Renewable Profile}\label{subsubsec:loadre}
The ADS uses 2018 hourly loads to project 2032 levels, resulting in values above current loads. We normalized these to current trends with a factor of 0.801, derived from CAISO’s DAM load forecasts for the CAISO area. These adjusted hourly profiles are used in the DAM simulation.

To create high-temporal-resolution load profiles for the RTM, we applied day-ahead forecast errors from 2022 CAISO data, mapped by TAC areas to our regions (PGE to Region 1, MWD/SCE/VEA to Region 2, SDGE to Region 3). Regional peak and average loads were aligned based on their TAC area distributions.

We used K-means clustering to apply historical day-ahead forecast errors to ADS hourly profiles, with Regions 4-7 derived using a linear regression model trained on DAM and RTM data from Regions 1-3. Wind and solar RTM profiles were generated similarly, using ADS data and CAISO predictions aggregated by region. Solar profiles were adjusted to zero for sunset periods and negative values. Final profiles reflect CAISO’s forecast error metrics, with mean absolute percentage errors of 1.7\% for load, 1.5\% for solar, and 6.5\% for wind~\cite{caisoforecast}.

\subsection{Two-Stage Settlement Model}
This paper models the wholesale electricity market as a two-stage settlement comprising an hourly-resolution DAM and a high-temporal-resolution RTM. Both can be modeled in look-ahead rolling horizon fashion.  In this study, supply-side market participants, encompassing owners of thermal generators, renewable generators, and emerging technologies, submit bids to DAM and RTM. For simplicity, we assume that thermal generators and renewable generators do not exercise market power. Accordingly, they bid at their actual marginal generation cost. Energy storage follows a distinct pricing principle. As a result, we utilize the CAISO 2022 storage hourly average day-ahead and real-time bids by quarter for energy storage bids~\cite{caisobattery}. We assume electricity demand to be inelastic, setting a high value of lost load (VOLL) to avoid infeasibility.

The market operator conducts a daily DAM to determine next-day unit commitments and schedules the dispatch for each generation of resources with day-ahead load and renewable generation forecast. The objective function of the DAM is to minimize system generation cost, subject to system and generation resources constraints. The constraints of DAM UC include:
\begin{itemize}
    \item Load balance constraints;
    \item System-wide operation reserve requirement constraints;
    \item Transmission line power flow limits;
    \item Thermal generator total and segment capacity, ramping, minimum up/down time, must run, and state transition constraints;
    \item Renewable generator availability constraints;
    \item Energy storage power rating, energy capacity, and state-of-charge evolution constraints.
\end{itemize}

Once the DAM UC is solved, the unit commitment outcomes are carried over to the RTM ED. Our test system enables the incorporation of a UC with a look-ahead window beyond the conventional 24-hour UC, aiming to better manage load and renewable fluctuations beyond the immediate operational day. On the operation day, high-temporal-resolution RTM ED is implemented in a multi-interval rolling horizon setting. With multi-period ED, only the first time interval decisions will be applied in the real dispatch. 

We implement the model in Julia, boasting high computational performance. Annual simulations can be readily executed on a standard personal laptop. The Julia implementation is available on GitHub\footnote{\url{https://github.com/niklauskun/STESTS}}. Detailed formulation for DAM UC and RTM ED is in Appendix.


\begin{figure}
    \centering
    \includegraphics[width=0.95\linewidth]{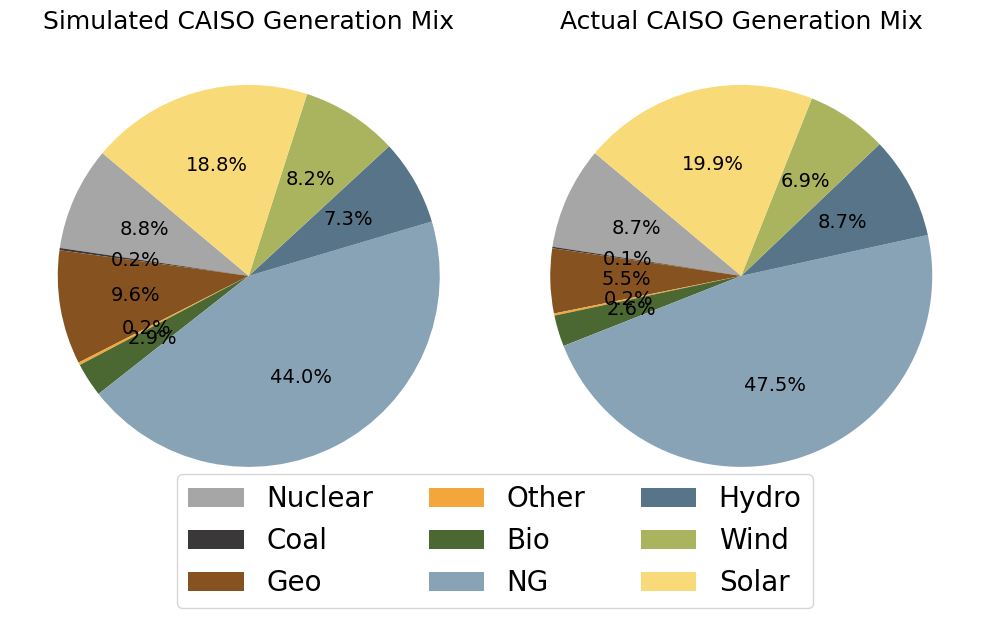}
    \caption{Simulated and 2022 CAISO historical year-long generation mix.}
    \label{fig:CAISO gen mix}
    \vspace{-1em}
\end{figure}

\begin{figure*}
\centering
    \subfigure[Winter peak week Feb 16-22.]{\includegraphics[trim = 0mm 0mm 0mm 0mm, clip, width = 1.55\columnwidth]{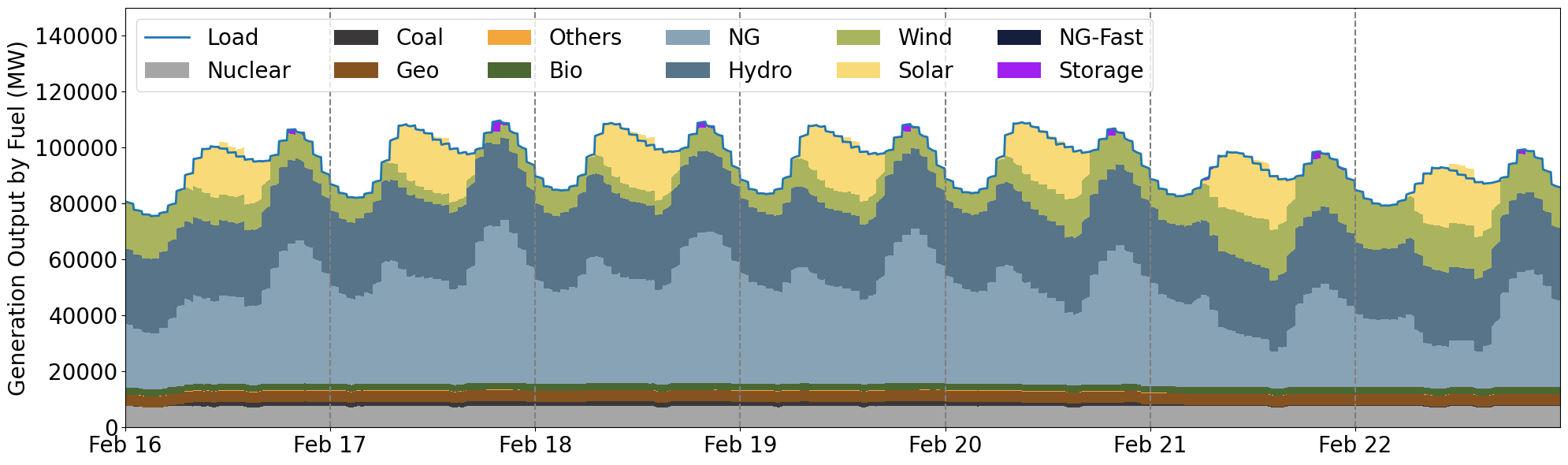}\label{winterpeak}}
    \hfill
    \subfigure[Summer peak week July 20-26.]{\includegraphics[trim = 0mm 0mm 0mm 0mm, clip, width = 1.55\columnwidth]{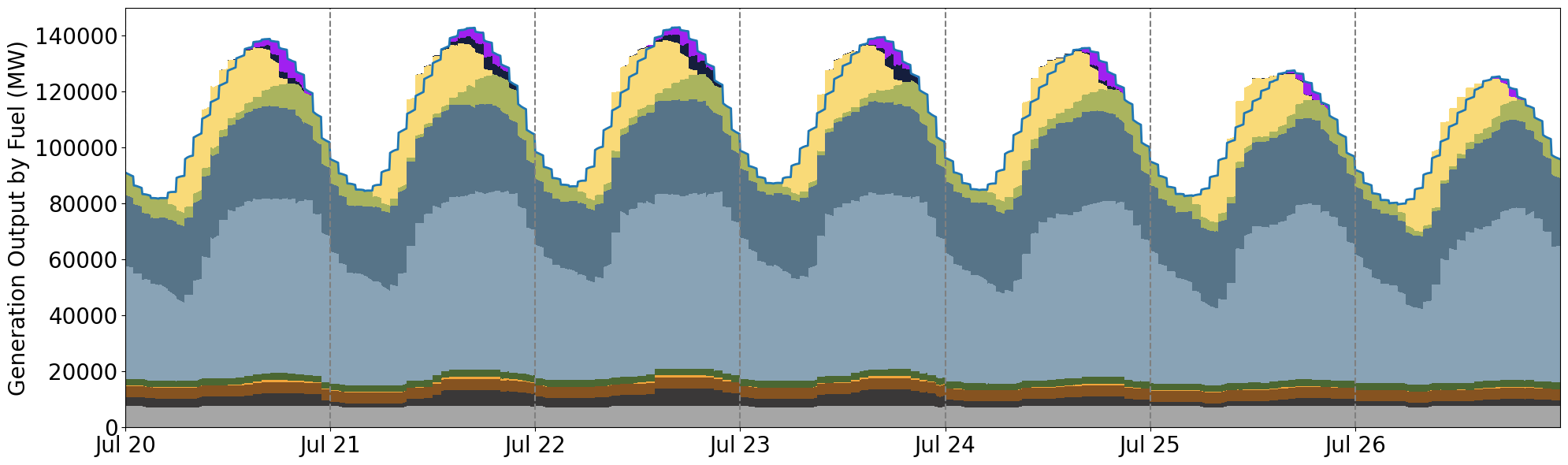}\label{fig.summerpeak}}
\caption{Total system generation by energy resources during winter and summer peak weeks. Two subfigures shared the same legend.}\label{fig*:weekgenmix}
\vspace{-1em}
\end{figure*}

\section{Analysis of Market Simulation Results}
We compare the wholesale market simulation results to historical data from CAISO, focusing on generation and price patterns. To ensure sufficient generation capacity for real-time dispatch, we set a 3\% operating reserve margin of the hourly predicted load in the DAM UC. Given that we simulate an energy-only market, the VOLL is set at \$9000/MWh. We assume that all energy storage units in the system submit identical bids for charging and discharging. These bids are derived from CAISO hourly average bids for the year 2022. Although we assume uniform bids in this case study, our test system offers a bidding interface, enabling each energy storage unit to submit strategic bids based on state-of-the-art bidding algorithms. Our year-long simulations use DAM UC without a look-ahead horizon and RTM ED with a thirteen 5-minute interval. For multi-interval rolling horizon implementation in DAM and RTM, we consider the solutions of the first 24 intervals (24 hours) in DAM and the first interval (5 minutes) in RTM as the actual cleared results. We also conduct a sensitivity analysis to explore the effects of varying look-ahead horizons in DAM and RTM.

\subsection{Generation by Energy Resources}

In this subsection, we show the generation pattern observed in our simulation result and compare it with historical statistics from CAISO. As highlighted in the data processing section, the renewable resources capacity in our test system closely mirrors the actual installed capacity in CAISO as of 2022. Fig.~\ref{fig:CAISO gen mix} compares annual generation mix by energy resources, where the simulated result is from the aggregation of Region 1 to 3, and the actual CAISO data is from California Energy Commission~\cite{CEC2022MIX}. The simulation generally aligns with the actual generation mix. 

Fig.~\ref{fig*:weekgenmix} displays the entire system generation by energy resources in real-time during the winter and summer peak weeks. Generations that exceed load represent energy storage charging power. We see pronounced seasonal patterns in load and renewable generations. Specifically, summer experiences higher loads, with a single peak occurring late afternoon. At the same time, winter sees relatively lower loads and two peaks - one in the morning and another in the evening. Coal power plants are seldom dispatched during winter peak week. However, they are committed to accommodating the overall higher demand in the summer. Both gas turbines and energy storage systems act as flexible resources in peak weeks. They are pivotal in addressing daily load variations and fluctuations within the hour. Energy storage exhibits higher utilization rates in the summer than in winter, often being charged during midday when solar energy peaks and discharged in the late afternoon to meet the surge in energy demand. Fast-start natural gas units are dispatched on the highest peak load days during ramping periods.

\subsection{Real-time Price}

\begin{figure}
\centering
\includegraphics[trim = 0mm 4mm 0mm 0mm, clip, width = 0.8\columnwidth]{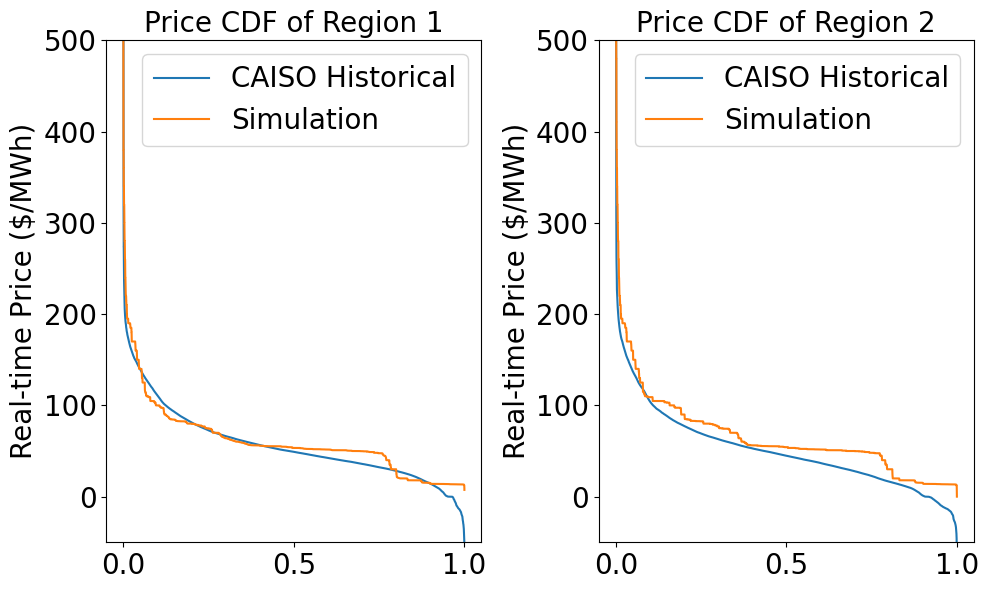}
\caption{Year-long real-time price cumulative density function comparison.}\label{fig:pricecdf}
\vspace{-1em}
\end{figure}

To further validate our test system, we compare the simulated real-time price patterns to the historical price distribution sourced from CAISO. The geographical locations of Region 1 and 2 is close to CAISO's price hubs NP15 and SP 15, respectively. Thus, Fig.~\ref{fig:pricecdf} compares the cumulative distribution function of the year-long simulated real-time prices and the 2022 real-time price data from CAISO hubs. Notably, the simulated real-time price duration curve is similar to historical data. This consistency is crucial, given that emerging technologies, such as energy storage and demand response operations, are sensitive to pricing.

However, there are non-smooth intervals observed within the simulated price duration curves. One reason for this is our assumption that all energy storage units submit identical bids at the hourly average historical bids by quarter. Additionally, we assume thermal generators and renewable resources bid truthfully at their marginal cost. We also notice short right tails in the simulation result, which indicates less low prices. Due to data limitations, our test system uses a reduced seven regions instead of an actual comprehensive network. This simplification leads to less congestion, which is one of the most common factors for negative prices within CAISO.

\begin{figure}
    \centering
    \subfigure[1 week example.]{\includegraphics[trim = 0mm 3mm 0mm 2mm, clip, width = 0.8\columnwidth]{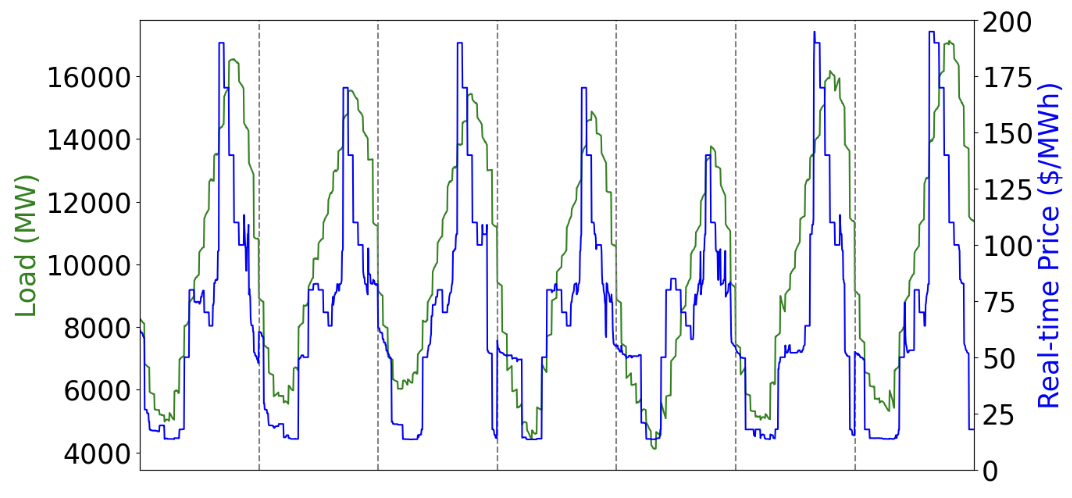}\label{fig.1wprice}}
    \hfill
    \subfigure[Monthly statistics.]{\includegraphics[trim = 5mm 5mm 0mm 5mm, clip, width = 0.8\columnwidth]{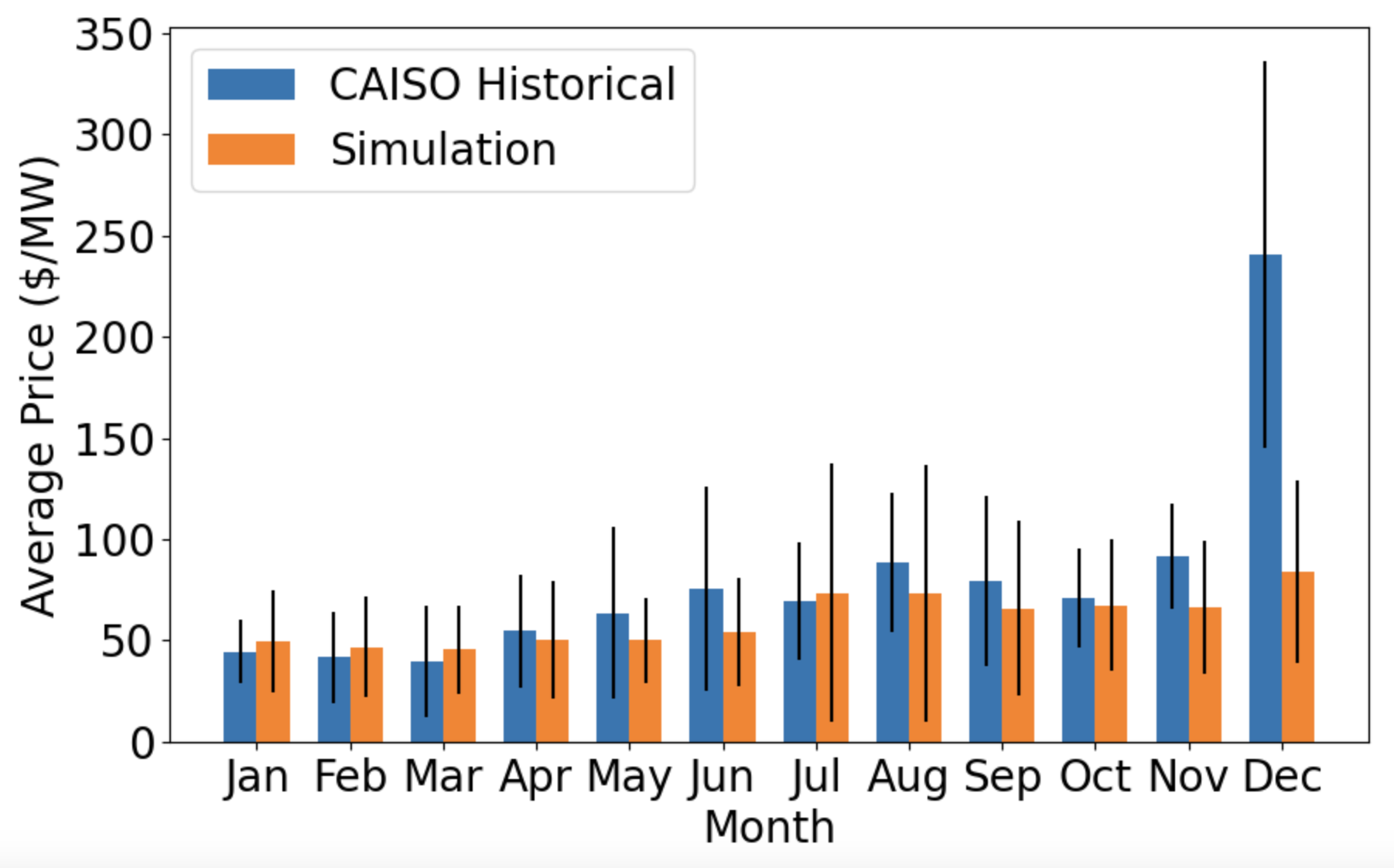}\label{fig.monthavg}}
    \caption{Example of simulated real-time price. (a) 1 week real-time price and load; (b) monthly statistics compared to CAISO historical data.}
    \label{fig:example}
    \vspace{-1em}
\end{figure}

Fig.~\ref{fig.1wprice} offers a snapshot of the real-time load and price over a week. 
We observe low prices in mid of the day and price surges during ramping-up periods. Figure~\ref{fig.monthavg} displays a month-by-month comparison of average real-time prices between CAISO historical data and market simulation model results, with error bars representing plus/minus one standard deviation. The patterns align closely for most of the year, indicating a strong correlation in monthly price trends. However, notable deviations emerge in November and December, where the simulated prices are significantly lower than historical prices. This divergence is attributed to an anomalous spike in natural gas prices from late November 2022 through January 2023, which is not captured by the simulation model.


\subsection{Sensitivity Analysis}
To manage multi-day and intra-hour uncertainties in renewables and storage, some operators use look-ahead windows with a rolling horizon in DAM and RTM. We ran a year-long simulation with varying ED and UC horizons to evaluate their impact. For ED, we tested look-ahead horizons of 1, 6, and 13 five-minute intervals, with the 13-interval case aligning with CAISO’s real-time practices. Including 6 and 13 intervals reduced system costs by 0.04\% and 0.12\%, respectively, improving intra-hour predictions of load and renewable changes. For UC, we examined 24-, 48-, and 72-hour look-ahead horizons. These showed minimal cost impact, as fast-response units in RTM balance the benefits of extended UC horizons. Cost savings may be slightly understated, as our model does not include contingencies or planned outages.

\section{Conclusion}
This study set a robust foundation for wholesale market simulation, designed to support detailed electricity market analysis, enabling detailed electricity market analysis with fine time granularity crucial for evaluating real-time market designs and integrating emerging energy resources. Initially tailored to CAISO, our WECC-based model closely reflects CAISO’s generation and pricing dynamics and can be adapted to other ISOs/RTOs with two-settlement markets using available generation, load, renewable profiles, and network data. Future research may enhance the model by integrating strategic bidding algorithms for emerging technologies and incorporating a more detailed network topology, supporting innovative market designs alongside the capabilities of new technologies.


\bibliographystyle{IEEEtran}	
\bibliography{IEEEabrv,main}		

\appendix

\subsection{Nomenclature}

\subsubsection{Sets and Indices}

\begin{IEEEdescription}[\IEEEusemathlabelsep\IEEEsetlabelwidth{$V_1,V_2,V_3$}]
\item[$Z$] Set of regions, indexed by $z$
\item[$G, G_z, G_m$] Set of thermal generators, subset of thermal generators in region $z$, and subset of must-run generators, indexed by $g$
\item[$GS_g$] Set of segments of thermal generator $g$, indexed by $gs$
\item[$GR, GR_z$] Set of renewable generators and subset of renewable generators in region $z$, indexed by $gr$
\item[$S, S_z$] Set of energy storage and subset of energy storage in region $z$, indexed by $s$
\item[$L, LT_z, LF_z$] Set of transmission lines and subset of transmission line to/from region $z$, indexed by $l$ 
\item[$H$] Set of hours for DAM, indexed by $h$
\item[$T$] Set of time intervals for RTM, indexed by $t$
\end{IEEEdescription}

\subsubsection{Variables}
\begin{IEEEdescription}[\IEEEusemathlabelsep\IEEEsetlabelwidth{$V_1,V_2,V_3$}]
\item[$gt_{g,h}$] Thermal generator output power, MW
\item[$gts_{g,gs,h}$] Thermal generator segment output power, MW
\item[$r_{g,h}$] Ramping product by thermal generator, MW
\item[$u_{g,h}$] Thermal generator commitment status, binary
\item[$v_{g,h}, w_{g,h}$] Thermal generator start-up/shut-down decision, binary
\item[$gr_{g,h}$] Renewable generator output power, MW
\item[$d_{s,h}, c_{s,h}$] Energy storage energy discharge/charge power, MW
\item[$f_{l,h}$] Line power flow, MW
\item[$\theta_{z,t}$] Voltage phase angle
\item[$s_{z,h}$] unserved energy, MW
\end{IEEEdescription}

\subsubsection{Parameters}
\begin{IEEEdescription}[\IEEEusemathlabelsep\IEEEsetlabelwidth{$V_1,V_2,V_3$}]
\item[$\mathrm{MC}_{g,gs}$] Generator segment marginal fuel cost, \$/MW
\item[$\mathrm{VOM}_{g}$] Generator variable operations and maintenance cost, \$/MW
\item[$\mathrm{NLC}_{g}$] Generator no-load cost, \$
\item[$\mathrm{SUC}_{g}$] Generator start-up cost, \$
\item[$\mathrm{P}^{max}_g,\mathrm{P}^{min}_g$] Generator maximum and minimum power output, MW
\item[$\mathrm{PS}^{max}_{g,gs}$] Generator segment maximum power output, MW
\item[$\mathrm{RU}_g,\mathrm{RU}_g$] Generator maximum ramp up/down rate, MW/hour
\item[$\mathrm{UT}_g, \mathrm{DT}_g$] Generator minimum up/down time, hour
\item[$\mathrm{SU}_g, \mathrm{SD}_g$] Number of hours generator must stay up/down since the start of the optimization horizon, hour
\item[$\mathrm{RA}_{gr,h}$] Renewable maximum expected capacity, MW
\item[$\mathrm{D}^{max}_s,\mathrm{D}^{min}_s$] Energy storage maximum and minimum discharge power rating, MW
\item[$\mathrm{C}^{max}_s,\mathrm{C}^{min}_s$] Energy storage maximum and minimum charge power rating, MW
\item[$\mathrm{\eta}^{D}_s,\mathrm{\eta}^{C}_s$] Energy storage discharge and charge efficiency
\item[$\mathrm{E}^{max}_s,\mathrm{E}^{min}_s$] Energy storage maximum and minimum energy capacity, MWh
\item[$\mathrm{E}^T_s$] Energy storage targeted state-of-charge at end of the operation day, MWh
\item[$\mathrm{B}^{D}_s,\mathrm{B}^{C}_s$] Energy storage discharge and charge bids, \$/MW
\item[$\mathrm{F}^{max}_l,\mathrm{F}^{min}_l$] Transmission line maximum and minimum power flow, MW
\item[$T(l),F(l)$] Transmission line $l$ to and from zone
\item[$\mathrm{X}_l$] Transmission line reactance, p.u.
\item[$\mathrm{L}_{z,h}$] Inflexible load, MW
\item[$\mathrm{RM}$] Opearation reserve marging
\item[$\mathrm{VOLL}$] Value of loss load, \$/MW
\item[$\mathrm{SC}$] Time scaling constant for RTM
\end{IEEEdescription}

\subsection{DAM UC Formulation}\label{app:UC}
\begin{subequations}
The objective of the DAM UC is to minimize the total system cost for the upcoming operational day. When incorporating a look-ahead horizon into the model, the optimization aims to minimize costs over the entire optimization period. This look-ahead feature allows the model to consider not just the immediate next day but also subsequent periods, providing a more holistic approach to system operation by anticipating future events and making decisions accordingly.
\begin{align}
\begin{gathered}
min \sum_{h}^{H} \Bigl[\sum_{g \in G} \Bigl(\sum_{gs \in GS} ({\mathrm{MC}_{g,gs}*gts_{g,gs,h})} 
\\+ VOM_g * gt_{g,h}+\mathrm{NLC}_g*u_{g,h}+\mathrm{SUC}_g*v_{g,h}\Bigl)
\\ \sum_{s}^{S} \Bigl(\mathrm{B}^D_{s,h} * d_{s,h} - \mathrm{B}^C_{s,h} * c_{s,h}\Bigl) + \sum_{z}^{Z} \mathrm{VOLL} * s_{z,h}\Bigl]
\end{gathered}
\end{align}

\subsubsection{System Level Constraints}
\begin{align}
\sum_{g \in G_z}gt_{g,h} &+ \sum_{gr \in GR_z}gr_{gr,h} +\sum_{s \in S_z} \left(d_{s,h}-c_{s,h}\right) \nonumber\\
+ \sum_{l \in LT_z}f_{l,h}& - \sum_{l \in LF_z}f_{l,h} + s_{z,h} = \mathrm{L}_{z,h} \label{cons:loadbalance}\\
\sum_{g \in G} &r_{g,h} \geq \mathrm{RM} \cdot \sum_{z \in Z} \mathrm{L}_{z,h} \label{cons:reserve}
\end{align}

System level constraints include load balance~\eqref{cons:loadbalance} and system operation reserve requirement~\eqref{cons:reserve}. The operation reserve requirement in our model is for whole system and each hour, proportional to the hourly load forecast.

\subsubsection{Transmission Constraints}
\begin{align}
\mathrm{F}^{min}_{l} &\leq f_{l,h} \leq \mathrm{F}^{max}_{l}\label{cons:transcap}\\
f_{l,h} &= \mathrm{X}_{l}*(\theta_{T(l),h}-\theta_{F(l),h})\label{cons:volt}\\
\theta_{ref,h} &= 0\label{cons:voltref}
\end{align}
We model transmission using DC-OPF. \eqref{cons:transcap} models the transmission capacity.  \eqref{cons:volt} and \eqref{cons:voltref} models the voltage phase angle, where $F(l)$, $T(l)$ are from region and to region of line $l$. $ref$ is the reference region.

\subsubsection{Thermal Generator Constraints}

\begin{align}
gt_{g,h} &= \mathrm{P}^{min}_g * u_{g,h} + \sum_{g \in G} gts_{g,gs,h} \label{cons:totalgen}\\
\mathrm{P}^{min}_g * u_{g,h} &\leq gt_{g,h} \leq \mathrm{P}^{max}_g * u_{g,h} \label{cons:gencap}\\
0 &\leq gts_{g,gs,h} \leq \mathrm{PS}^{max}_{g,gs} \label{cons:segcap}\\
-gt_{g,h} + gt_{g,h-1} &\leq \mathrm{RD}_g + \mathrm{P}^{min}_g * w_{g,h} \label{cons:rampdown}\\
gt_{g,h} - gt_{g,h-1} &\leq \mathrm{RU}_g + \mathrm{P}^{min}_g * v_{g,h} \label{cons:rampup} \\
r_{g,h} &\leq \mathrm{P}^{max}_{g} \cdot u_{g,h} - gt_{g,h} \label{cons:ramp1}\\
r_{g,h} &\leq \mathrm{RU}_{g}\label{cons:ramp2}\\
u_{g,h} &= 1, \quad \forall g \in G_m \label{cons:mustrun}\\
v_{g,h} - w_{g,h} &= u_{g,h} - u_{g,h-1}\label{cons:state}\\
v_{g,h} + w_{g,h} &\leq 1\\
\sum_{\tau = \max\{h - \mathrm{UT}_g + 1, 1\}}^h v_{g,\tau} &\leq u_{g,h},  h\in\{\mathrm{SU}_g,\dotsc, H\}\\
\sum_{\tau = \max\{h - \mathrm{DT}_g + 1,1\}}^h w_{g,\tau} &\leq 1-u_{g,h}, h\in\{\mathrm{SD}_g,\dotsc H\}\label{cons:mustdown}
\end{align}

\eqref{cons:totalgen}, \eqref{cons:gencap}, and \eqref{cons:segcap} models the total output and segment output of a thermal generator. \eqref{cons:rampdown} and \eqref{cons:rampup} are ramp-up and ramp-down constraints. At the beginning of optimization horizon, $gt_{g,h-1}$ in \eqref{cons:rampdown} and \eqref{cons:rampup} are parameters representing initial output. \eqref{cons:ramp1} and \eqref{cons:ramp2} limit the ability of generators providing operation ramping products. \eqref{cons:mustrun} is the must-run constraints. \eqref{cons:state}--\eqref{cons:mustdown} are state transition and must up and down time constraints. With a rolling horizon, the remaining minimum up and down time at the end of operation day will be passed to next DAM UC.

\subsubsection{Renewable Generator Availability Constraints}
\begin{equation}
0 \leq gr_{gr,h} \leq \mathrm{RA}_{gr,h} \label{cons:re}
\end{equation}

Due to the limitation of the hydro power system data, hydro resources are similarly modeled as solar and wind resources. The maximum generation from each renewable generation unit may not exceed the maximum expected capacity. 

\subsubsection{Energy Storage Constraints}

\begin{align}
\mathrm{D}^{min}_{s} &\leq d_{s,h} \leq \mathrm{D}^{max}_{s} \label{cons:drate}\\
\mathrm{C}^{min}_{s} &\leq c_{s,h} \leq \mathrm{C}^{max}_{s} \label{cons:crate}\\
e_{s,h} - e_{s,h-1} &= \mathrm{\eta}^C_s * c_{s,h}- \frac{1}{\mathrm{\eta}^D_s} *d_{s,h} \label{cons:SOC}\\
\mathrm{E}^{min}_{s} &\leq e_{s,h} \leq \mathrm{E}^{max}_{s} \label{cons:energycap}\\
e_{s,t} \geq E^T_s,& \quad \forall \{ t \in T \mid t \equiv 0 \ (\text{mod} \ 24) \} \label{cons:endsoc}
\end{align}

Constraints for energy storage systems include discharge and charge power capacity~\eqref{cons:drate},\eqref{cons:crate}, state-of-charge evolution~\eqref{cons:SOC}, and energy capacity~\eqref{cons:energycap}. At the begining of optimization horizon, $e_{t-1}$ in \eqref{cons:SOC} is a parameter representing initial state-of-charge. The targeted state-of-charge is defined by~\eqref{cons:endsoc}.

\end{subequations}

\subsection{RTM ED Formulation}\label{app:ED}
In the context of a rolling-horizon with a look-ahead window, the RTM ED borrows certain similarities from the DAM UC formulation. In the DAM UC, unit commitment decisions are made using binary variables $u_{g,h}$, $v_{g,h}$, and $w_{g,h}$. These decisions, when transitioned to the RTM ED context, transform into parameters, denoted as $U_{g,t}$, $V_{g,t}$, and $W_{g,t}$. Here's a breakdown of their correspondence:
\begin{itemize}
    \item $U_{g,t}$ in the RTM ED represents whether a unit is committed during a particular hour. It will be set to 1 if its corresponding hour variable $u_{g,h}$ is 1.
    \item $V_{g,t}$ in the RTM ED indicates the start-up decision of a unit at the beginning of an hour interval. It will be set to 1 only if $t$ is interval at beginning of an operation hour and $v_{g,h}$ in the DAM UC for that hour is 1.
    \item $W_{g,t}$ in the RTM ED indicates the shut-down decision of a unit at the beginning of an hour interval. It will be set to 1 only if $t$ is interval at beginning of an operation hour and $w_{g,h}$ in the DAM UC for that hour is 1.
\end{itemize}

The objective function of RTM ED is
\begin{subequations}
\begin{align}
\begin{gathered}
\min \quad SC*\sum_{t}^{T} \Bigl[\sum_{g \in G} \Bigl(\sum_{gs \in GS} ({\mathrm{MC}_{g,gs}*gts_{g,gs,t})} 
\\+ VOM_g * gt_{g,t}+\mathrm{NLC}_g*U_{g,t}+\mathrm{SUC}_g*V_{g,t}\Bigl)
\\ \sum_{s}^{S} \Bigl(\mathrm{B}^D_{s,t} * d_{s,t} - \mathrm{B}^C_{s,t} * c_{s,t}\Bigl) + \sum_{z}^{Z} \mathrm{VOLL} * s_{z,t}\Bigl]
\end{gathered}
\end{align}

Subject to constraints include load balance constraint~\eqref{cons:loadbalance}, transmission constraints~\eqref{cons:transcap}--\eqref{cons:voltref}, thermal generator constraints~\eqref{cons:totalgen}--\eqref{cons:rampup}, renewable availability constraint~\eqref{cons:re}, and energy storage constraints~\eqref{cons:drate}--\eqref{cons:energycap}. All variables with time interval indexed by $h$ are replaced with shorter interval indexed $t$.

Note that energy storage state-of-charge evolution constraint~\eqref{cons:SOC} and ramping constraints~\eqref{cons:rampdown},\eqref{cons:rampup} are scaled by time scaling constant $\mathrm{SC}$:

\begin{align}
e_{s,t} - e_{s,t-1} &= \mathrm{SC}*(\mathrm{\eta}^C_s * c_{s,t}- \frac{1}{\mathrm{\eta}^D_s} *d_{s,t})\\
-gt_{g,t} + gt_{g,t-1} &\leq \mathrm{SC}*\mathrm{RD}_g + \mathrm{P}^{min}_g * W_{g,t}\\
gt_{g,t} - gt_{g,t-1} &\leq \mathrm{SC}*\mathrm{RU}_g + \mathrm{P}^{min}_g * V_{g,t}
\end{align}
\end{subequations}

\end{document}